# The Economic Trend of Video Game Industry

**[1] Guanxi Zhuang, [2]Hai Zhang, [3]Xia Liu**

[*1]University of Science and Technology of China
[2]Beijing University of Posts and Telecommunications, Computer Science School, Beijing, China

**Abstract** –In recent years the game industry has had a huge growth. We've seen new game consoles, great looking games and an increase in the number of people playing them. We are presently in the seventh generation of video games which focuses on consoles released since 2004. For home consoles, the seventh generation began on November 22, 2005 with the release of Xbox 360 and continued with the release of PlayStation 3 on November 11, 2006, and Wii on November 19, 2006. The current generation is having a console battle between Nintendo's Wii, Microsoft's Xbox 360, and Sony's PlayStation 3. The appearance of the three new consoles not only offers various purchase choices, but also greatly affects economy and culture.

**Keywords** –PS3; PlayStation; Xbox 360

## 1. Purchase puzzle

A person buying a console will surely get baffled in deciding between these three. Each new console has introduced a new type of breakthrough in technology. The Xbox 360 offered games rendered natively at High Definition (HD) resolutions. In addition to HD gaming, the PlayStation 3 offered HD movie playback via a built-in Blu-ray Disc player. The Wii, however, created an entirely new gaming way, which focused on integrating controllers with movement sensors as well as joysticks. One thing to keep in mind is that every machine has its own plus and minus points. You need to take your decision based on how you want to use the console.

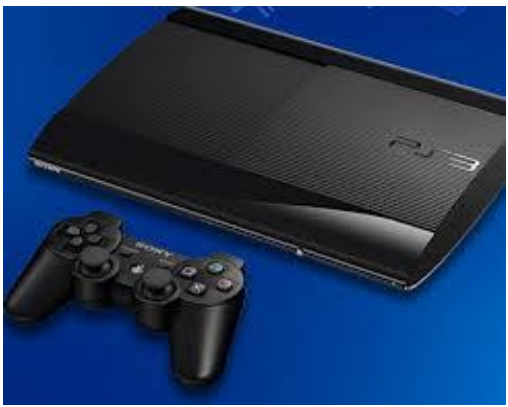

**Figure 1.** The photo of PS 3

If you are a casual gamer or want to buy a game console for your kids, you can consider the Wii. The Wii is no doubt a great console and is in fact pretty much ahead of the other two in terms of sales. It has the advantage of bringing an entirely new gaming experience. It utilizes motions based on player movements to interact with characters on-screen. With Wii, users are able to physically move their bodies while playing a game. But you need beware that the Wii's visuals or graphics will not be comparable with the Xbox 360 or the PS3 as it doesn't have powerful hardware backing it.

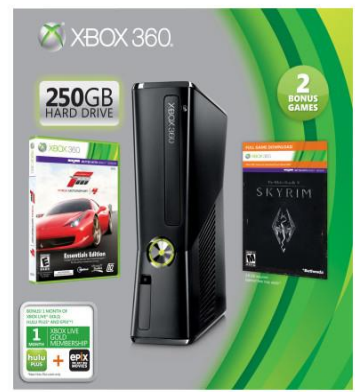

**Figure 2.** the photo of Xbox 360

If you are a hardcore gamer, the PS3 and the Xbox 360 are good choices. Both consoles are quite powerful and have strong hardware supporting them. They also offer superb visuals and games including the best exclusive games. Considering the price, the new PS3 Slim and Xbox 360 Elite both go for $299. If you want to use your console for both gaming and media purposes, then go for the PS3. It has a Blu-Ray drive and the built-in WiFi makes streaming media easy. If you are a First-

Person Shooter (FPS) game lover, the 360 can be good as it has a lot of shooter games.

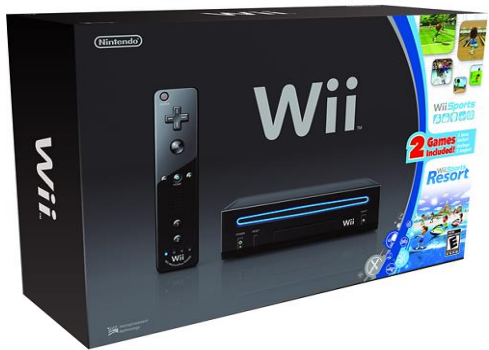

**Figure 3.** the shape of Wii

## 2. Economic impact

The video game industry is one of the fastest growing sectors in the U.S. economy as it is stated by The Entertainment Software Association:

"From 2005 to 2009, the industry's real rate of growth was more than seven times the real rate of growth for the entire economy. In addition, computer and video game companies posted strong overall sales in 2011 with revenues of nearly $25 billion as entertainment software companies continue to provide jobs to state and local economies across the nation." ("Games: Improving the Economy")

The appearance of the Wii, the Xbox 360 and the PS3 in the last five years contributes significantly to the games industry's economic growth as well as promotes the development of the game software industry by attracting more and more games to be created on different platforms.

In 2009, U.S. retail sales of computer and video games reached nearly $11 billion. Sales figures from each console's launch date were added up, with the Wii having the top sale and 360 just barely edging out the PS3: 96.7 million for the Wii, 68.6 million for the 360, and 66.8 million for the PlayStation 3. A recent study, "Video Games in the 21st Century: The 2010 Report" detailed the impact that computer and video game companies have on America's economy. The report stated:

"From 2005 to 2010, the entertainment software industry's revenue more than doubled. Over the same period, the entire U.S. Gross Domestic Product (GDP) only grew by about 16%. The entertainment software industry added nearly $5 billion to the U.S. GDP in 2009." (Siwek 3)

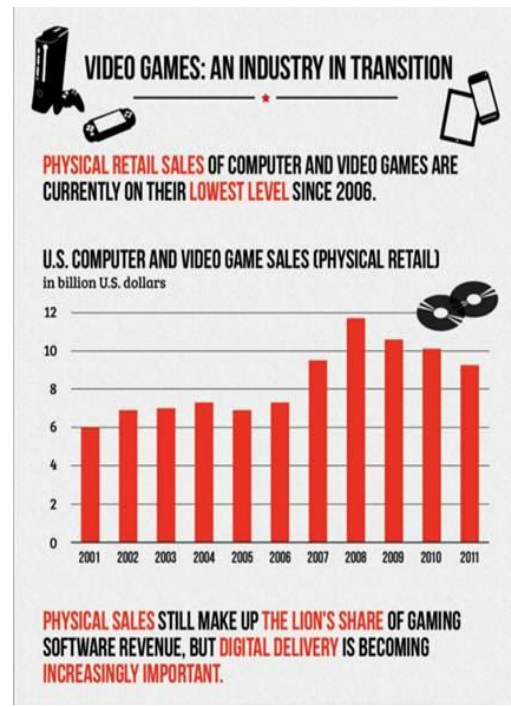

**Figure 1.** retail sales of computer and video games from 2006.

Also more game jobs are offered by many entertainment software companies. Several states in the U.S. including California, New York and Washington have the highest number of video game jobs. Over 22,000 workers are employed in these areas. This number is almost 71 percent of the industry's total direct employment.

## 3. Culture influence

With the popularity of the game consoles, US authorities begin to fear that criminals like terrorists are using games consoles to interact with each other. In order to find a way to capture the data from popular consoles such as the Xbox 360, the Wii and the PS3, the US navy assigned a £11,000 contract to Obscure Technologies Company to develop a prototype rig.

Besides the potential criminals, the PS3 game console is also involved in the military research. The US Air Force Research Laboratory's (AFRL) took advantages of the PS3's powerful hardware and the unique Linux system, and created the cheapest supercomputer by connecting 1,760 PS3 systems together. This supercomputer is used for research by Air Force Service branches including quick processing of ultra-high-resolution satellite imagery, as well as research into artificial intelligence, pattern recognition and radar enhancement.

In addition to the influence on the government, as to kids or the young generation, their life revolves around the Xbox, the PS3, and the Wii. Comparing to the Xbox and PS3, the Wii has much more child-oriented games that are cute and fun. Playing the Wii can be a new form of exercise. Sean's research stated that exercising regularly by playing Wii video games presents a potential psychological benefit. Children don't have to wear sporting shoes and step outside to exercise. The whole families can stay together and enjoy a fascinating, fun and immersive exercising atmosphere.